\documentclass[superscriptaddress,aps,arXiv,reprint,twocolumn,amsmath,amssymb]{revtex4-2}
\usepackage{graphicx}
\usepackage{hyperref}
\usepackage{amssymb}
\usepackage{slashed}
\usepackage{dcolumn}
\usepackage{amsmath}
\usepackage{bm}
\usepackage{colordvi}
\usepackage{algorithm}
\usepackage{algpseudocode}
\usepackage{multirow}
\usepackage{titlesec}
\usepackage{newtxtext,newtxmath}
\usepackage{dsfont}
\usepackage{xcolor}
\usepackage{mathbbol}

\usepackage{hyperref}
\hypersetup{
    colorlinks=true,
    linkcolor=blue,
    citecolor=blue,     
    urlcolor=blue,
}

\allowdisplaybreaks

\usepackage{mathrsfs}
\makeatletter

\newcommand{\Rmnum}[1]{\expandafter\@slowromancap\romannumeral #1@}
\makeatother

\begin{document}

\preprint{APS/123-QED}

\title{Correlation-Driven $d$-Wave Superconducting Dome from Pseudogap Spectral Reconstruction}


\author{Yue Yuan}
\email{yueyuanyy@mail.ustc.edu.cn}
\affiliation{Department of Modern Physics, University of Science and Technology of China, Hefei 230026, China}

\author{Ziwen Pan}
\email{panzw19@ustc.edu.cn}
\affiliation{Department of Modern Physics, University of Science and Technology of China, Hefei 230026, China}

\author{F. Yang}
\email{yfgq@ustc.edu.cn}
\affiliation{Department of Physics,  University of Science and Technology of China, Hefei,
Anhui, 230026, China}

 \date{\today}
 
\begin{abstract}
  Previous theoretical studies  [Nat. Phys. {\bf 16}, 1175 (2020)] based on the Hatsugai–Kohmoto model have examined the stability of $s$-wave  superconductivity in strongly correlated systems, demonstrating that correlations alone can substantially modify superconducting behavior.  Motivated by this perspective, but going beyond these studies, we perform self-consistent microscopic calculations of $d$-wave superconductivity in strongly correlated systems by employing an exactly solvable correlated model that hosts a pseudogap phase and a partially flat band [Phys. Rev. Lett. {\bf 133}, 166501 (2024)]. We show that pseudogap correlations and superconducting order affect the low-energy spectrum in qualitatively different ways: the former leads to a momentum-localized suppression of spectral weight, whereas the latter induces a coherent reorganization of quasiparticle excitations. Moreover, we  demonstrate that the interplay between superconducting order and pseudogap correlations naturally generates a superconducting dome in the temperature-doping phase diagram, with optimal doping located near the quantum critical point separating the pseudogap and metallic phases. Furthermore, $d_{x^2-y^2}$-wave superconductivity is found to be remarkably robust, remaining energetically dominant over both $d_{xy}$-wave and $s$-wave pairing channels across a wide doping range. Our results offer a potential route for a direct and controlled connection between pseudogap correlations and the emergence of the superconducting dome in cuprates.
\end{abstract}

\maketitle


{\sl Introduction.---}Over the past several decades, tremendous theoretical efforts have been devoted to understanding strong electronic correlations and high-temperature superconductivity in cuprate materials~\cite{bednorz86possible,wu1987superconductivity,keimer15from,dagotto94correlated,tom99the,armitage10progress,davis13concepts,PhysRevB.102.014511}. One of the central challenges is to account for the characteristic superconducting (SC) dome with $d_{x^2-y^2}$-wave pairing symmetry~\cite{keimer15from}, which peaks near optimal doping. Equally important is the nature of the normal state: on the underdoped side, a pseudogap phase emerges~\cite{Ding1996,loeser1996excitation,norman1998,gomes2007visualizing,pasupathy2008electronic,alldredge2008evolution}, characterized by a finite excitation gap persisting above the SC  transition temperature $T_c$ up to a much higher temperature $T^*$, whereas the overdoped regime exhibits strange-metal behavior~\cite{martin1990normal,keimer15from}, exhibiting transport behaviors that deviate from those of conventional metals.  A key insight from previous theoretical studies is that superconductivity emerging from an uncorrelated or weakly correlated normal state is generically smooth and featureless: the transition temperature $T_c$ varies monotonically with doping and fails to form a dome~\cite{li21superconductor,yang21theory}. This observation strongly suggests that the pseudogap and strange-metal phases, i.e., the correlated normal-state electronic structures, play an essential role in shaping superconductivity in cuprates.  Despite extensive efforts, achieving a unified theoretical description remains highly challenging. Numerical approaches based on the Hubbard model~\cite{annurev:/content/journals/10.1146/annurev-conmatphys-090921-033948,PhysRevLett.102.056404,PhysRevX.11.011058,PhysRevB.97.085125,PhysRevX.8.021048,RevModPhys.68.13,vollhardt2011dynamical,avella2013strongly} become computationally intractable in spatial dimensions larger than one, particularly when long-range correlations and competing orders are involved. 

This limitation has motivated the search for analytically tractable models that capture essential correlation effects beyond weak coupling. Recently, a growing body of work has explored the Hatsugai–Kohmoto (HK) model~\cite{phillips18absence,phillips20exact,yeo19local,zhao22thermodynamics,zhao23failure,zhong22solvable,zhu21topological,tenkila25dynamical,mai23is,mai23topological,setty24electronic,setty24symmetry,PhysRevB.108.235149,li2022two,PhysRevB.108.035121}, an exactly solvable variant of the Hubbard model that hosts a Mott insulator at half filling and a non-Fermi-liquid metallic state upon doping~\cite{phillips20exact,zhao22thermodynamics}. Building on this framework, Phillips {\em et al.}~\cite{phillips20exact} investigated superconductivity arising from the HK model and demonstrated that strong correlations alone can qualitatively modify the SC phase diagram. While these results are valuable, the HK model suffers from two fundamental limitations in the context of cuprates: it does not capture the pseudogap phase, and it naturally favors $s$-wave pairing rather than the experimentally established $d$-wave symmetry. Very recently, Worm {\em et al.}~\cite{worm24fermi} introduced an exactly solvable correlated model featuring a partially flat band structure and, crucially, a pseudogap phase separating a correlated metal from a Fermi-liquid regime. This development provides a new opportunity to revisit superconductivity in a setting that incorporates both strong correlations and a pseudogap at the level of the normal state. In direct contrast to   numerous previous studies based on the HK model~\cite{phillips18absence,phillips20exact,yeo19local,zhao22thermodynamics,zhao23failure,zhong22solvable,zhu21topological,tenkila25dynamical,mai23is,mai23topological,setty24electronic,setty24symmetry,PhysRevB.108.235149,li2022two,PhysRevB.108.035121}, it is therefore natural to investigate $d$-wave superconductivity in this model and to explore its interplay with the pseudogap correlations.

In this work, we perform such an analysis within a self-consistent microscopic calculation. We show that pseudogap correlations play a decisive role in shaping superconductivity and lead to several salient features that closely resemble experimental observations in cuprates. In particular, we show that pseudogap correlations and SC  order have qualitatively different impacts on the low-energy spectrum: the pseudogap causes an incoherent and momentum-localized depletion of spectral weight, whereas superconductivity induces a coherent reorganization of quasiparticle excitations. We find a SC dome in the $T$-$p$ phase diagram arising from the competition between SC order and the pseudogap, with the optimal doping located near the quantum critical point separating the pseudogap and metallic phases. These findings suggest the essential role of correlated normal-state physics in stabilizing high-$T_c$ superconductivity and provide an analytically controlled method for understanding the emergence of the SC dome in cuprates.

{\sl A Solvable Correlated Model.---}The solvable correlated model employed to describe the normal-state electronic structure was originally introduced by Worm et al.~\cite{worm24fermi}. There, the model was extensively benchmarked against numerically demanding many-body calculations for the full local Hubbard model using the dynamical vertex approximation (D$\Gamma$A), demonstrating that it faithfully captures the essential correlation effects in the doped regime. {Specifically, following Ref.~\cite{worm24fermi}, the effective normal-state Hamiltonian is given by 
\begin{equation}\label{Hamiltonian}
    \hat{H}=\!\!\sum_{\bm{k} \sigma} \Big(\epsilon_{\bm{k}} \hat{n}_{\bm{k} \sigma}+ \frac{{{\mathcal{V}}}}{2}\hat{n}_{\bm{k} \sigma} \hat{n}_{\bm{k} + \bm{Q},-{\sigma}} \Big), 
\end{equation}
where $\mathbf{Q}=(\pi,\pi)$, and $\hat{n}_{\bm{k} \sigma} = \hat{c}_{\bm{k} \sigma}^\dagger \hat{c}_{\bm{k} \sigma}$ is the density operator with $\hat{c}_{\bm{k} \sigma}$ ($\hat{c}_{\bm{k} \sigma}^\dagger$) being the annihilation (creation) operators of an electron with momentum $\bm{k}$ and spin $\sigma$; $\mathcal{V}$ denotes the effective interaction strengths associated with local correlations and antiferromagnetic-fluctuation-mediated interactions. The single-particle dispersion on a square lattice is taken as 
\begin{align}
\epsilon_{\bm{k}} =&-2 t [\cos(k_x) + \cos(k_y)] - 4 t' \cos(k_x) \cos(k_y)\nonumber\\
&- 2 t'' [\cos(2 k_x) + \cos(2 k_y)]-\mu,
\end{align}
where $\mu$ is the chemical potential. It is worth noting that, in contrast to the HK model with $\mathbf Q=0$, whose momentum-local interaction produces a Mott insulator at half filling and a non–Fermi-liquid metal upon doping, the solvable correlated model of Ref.~\cite{worm24fermi} adopts a finite ordering vector ${\bf Q}=(\pi,\pi)$, thereby coupling states at $\mathbf k$ and $\mathbf k+\mathbf Q$ and leading to a qualitatively different correlated structure. Physically, this restriction isolates the coupling mediated by antiferromagnetic (AFM) spin fluctuations, which constitute the leading low-energy interaction channel in cuprates. Importantly, Ref.~\cite{worm24fermi} demonstrated that this simplified interaction reproduces key features of the full Hubbard model obtained from D$\Gamma$A calculations, including the doping evolution of the Fermi surface and momentum-resolved spectral functions, particularly away from the parent Mott insulating phase. Thus, the HK-coupling and AFM-fluctuations-mediated interactions  correspond to distinct fixed-point structures governing different doping regimes. The remaining interaction channels primarily renormalize the effective strengths of these dominant couplings without altering their qualitative form. More detailed discussions can be found in Ref.~\cite{worm24fermi}.}

{In cuprates, the Mott-insulating phase is confined to a narrow region near $p=0$, 
and the system rapidly enters the pseudogap regime upon doping.}  Since the Mott insulating order does not coexist with superconductivity in cuprates~\cite{keimer15from}, we focus on the regime where local Mott localization effects are suppressed and set ${\bf Q}=(\pi,\pi)$. This allows us to isolate the interplay between superconductivity and the correlated metallic and pseudogap phases encoded by the AFM-fluctuation-mediated interaction $\mathcal{V}$.  In this regime, the Hamiltonian~(\ref{Hamiltonian}) remains exactly solvable.  In the occupation-number basis $|a,b\rangle$, where $a$ and $b$ denote the occupations of the $\bm{k}$ and $\bm{k}+\bm{Q}$ states, respectively, the block Hamiltonian $\hat{H}_{\bm{k}\sigma}$ becomes diagonal: 
\begin{equation}
    \hat{H}_{\bm{k}\sigma} = \mathrm{diag}\{ 0, \epsilon_{\bm{k}}, \epsilon_{\bm{k}+\bm{Q}}, \epsilon_{\bm{k}} + \epsilon_{\bm{k}+\bm{Q}} + {\mathcal{V}} \}.
\end{equation} 
and the single-particle excitation spectrum is given by~\cite{worm24fermi}
\begin{equation}
    A(\bm{k},\omega) = (1 - n_{\bm{k} + \bm{Q}}) \delta(\omega - \epsilon_{\bm{k}}) + n_{\bm{k} + \bm{Q}} \delta(\omega - \epsilon_{\bm{k}} - {{\mathcal{V}}}), \label{eq:A}
\end{equation}
where the electron occupation is given by
\begin{equation}
n_{\bm{k}+\bm{Q}}=\frac{e^{-\beta\epsilon_{{\bm{k}+\bm{Q}}}}+e^{-\beta(\epsilon_{\bm{k}}+\epsilon_{{\bm{k}+\bm{Q}}}+{{\mathcal{V}}})}}{1+e^{-\beta\epsilon_{\bm{k}}}+e^{-\beta\epsilon_{{\bm{k}+\bm{Q}}}}+e^{-\beta(\epsilon_{\bm{k}}+\epsilon_{{\bm{k}+\bm{Q}}}+{{\mathcal{V}}})}}.
\end{equation}
Here $\beta=1/(k_B T)$, with $k_B$ being the Boltzmann constant and $T$ being the temperature. 

All parameters are taken directly from Ref.~\cite{worm24fermi} without further adjustment. The nearest-, next-nearest-, and next-next-nearest-neighbor hopping amplitudes are chosen as
$t \approx 0.3\,\mathrm{eV}$, $t' \approx -0.2t$, $t'' \approx 0.1t$. Moreover, direct numerical comparisons between the solvable correlated model and the full Hubbard model have demonstrated that the former qualitatively reproduces the results of numerically demanding many-body calculations for the latter, and within this mapping, $\mathcal{V}=1.1t$ in the solvable correlated model corresponds to a Hubbard interaction strength $U=8t$ in the original Hubbard model~\cite{worm24fermi}. 

In the calculation, the doping level $p$ is defined through the electron density,
\begin{equation}
p = 1 - n, \qquad n = \frac{1}{N}\sum_{{\bf k}\sigma} \langle \hat{n}_{\bf{k}\sigma} \rangle,
\end{equation}
which is determined self-consistently by the chemical potential $\mu$ at a given temperature.  In practice, we vary $\mu$ to tune the filling $n$, and thereby obtain the corresponding hole concentration $p$. It should be emphasized in the doping range considered here, the relevant {\emph{ground state}} is the pseudogap state,  rather than a Mott-insulating state.  \textcolor{black}{
 As demonstrated in Ref.~\cite{worm24fermi}, the exactly solvable correlated model was extensively benchmarked against numerically demanding calculations for the Hubbard model based on the dynamical vertex approximation (D$\Gamma$A). In particular, it was shown to reproduce the essential doping evolution of the momentum-resolved spectral function, Fermi-surface reconstruction, and pseudogap phenomenology of the Hubbard model over a broad doping range away from the Mott-insulating limit. Therefore, the physical relevance of the present theory does not rely on a direct description of the Mott-insulating phase, but rather on its demonstrated ability to reproduce the correlated electronic structure of the pseudogap regime obtained from Hubbard-model calculations. Accordingly, in our simulations, we restrict $p$ to finite values  and $p$ serves as a physically meaningful filling parameter determined self-consistently by the chemical potential, which characterizes the evolution of the correlated pseudogap and metallic regimes described by the model. From this perspective, the comparison with phenomenology of cuprates is intended at the level of the phase-diagram evolution, spectral reconstruction, and superconducting dome formation, rather than as a quantitative mapping of the entire doping axis.}

\begin{figure}
    \centering
  {\includegraphics[width=8.4cm]{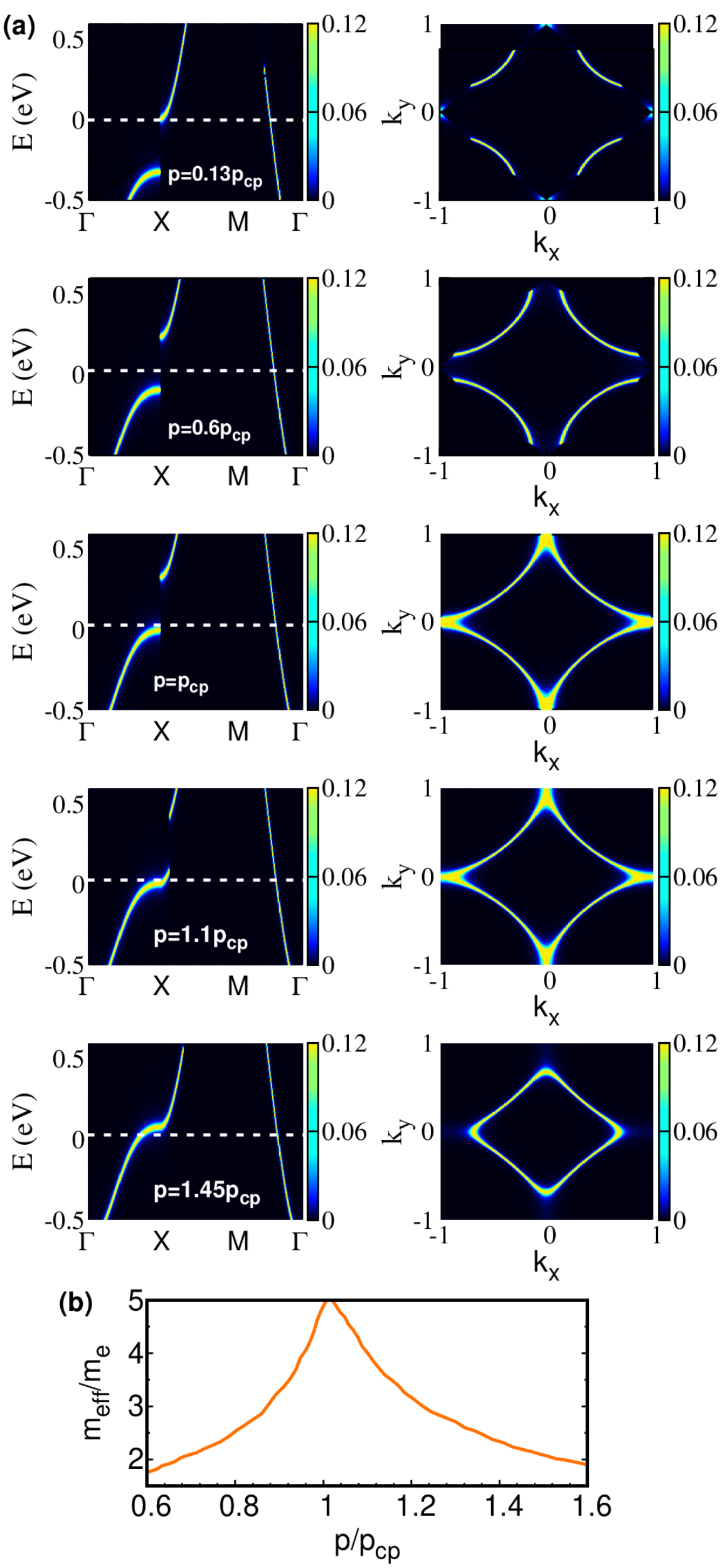}}
    \caption{(a) Electronic structures of the normal-state phase at $k_BT\approx10^{-4}t$ for underdoped, quantum-critical point of pseudogap phase and overdoped cases, obtained from the solvable correlated model.
    The Fermi level is located at $E=0$ (dashed line).
    Left column: Single-particle excitation spectra along a high-symmetry path $\Gamma=(0, 0)$, $X=(\pi, 0)$, and $M=(\pi, \pi)$ in the Brillouin zone. Right column: Corresponding Fermi surfaces.
    (b) Effective mass (i.e., normalized DOS) averaged on Fermi surface, incorporating the correlation effect.}
    \label{fig:normal}
\end{figure}

The normal-state spectral function at a specific doping level (underdoped, $p=0.43p_{\rm cp}$) was discussed in Ref.~~\cite{worm24fermi}. Here, we provide a systematic study of the doping dependence over a broad doping range, which serves as a foundation for the subsequent investigation of superconductivity.  Figure~\ref{fig:normal}(a) summarizes the hole-doping evolution of the single-particle excitation spectrum and Fermi surface topology as a function.
 At low doping ($p<p_{\rm cp}$, with $p_{\rm cp}$ being the quantum critical point of the pseudogap phase), corresponding to the pseudogap regime, strong correlation effects lead to a suppression of low-energy spectral weight in the antinodal region near ${\bm k}=(\pi,0)$. The chemical potential lies inside this correlation-induced gap, eliminating the conventional Fermi surface in this momentum sector. Instead, a Luttinger surface characterizes the low-energy structure~\cite{worm24fermi,PhysRevB.68.085113}. Moving away from the antinodal region, coherent quasiparticle states re-emerge, resulting in disconnected segments of Fermi surface commonly referred to as Fermi arcs. This momentum-selective reconstruction closely mirrors angle-resolved photoemission spectroscopy observations in underdoped cuprates~\cite{RevModPhys.75.473,Vishik_2018}.

{From the spectral function $A({\bf k},\omega)$ [Eq.~(\ref{eq:A})], the pseudogap arises from the correlation-induced splitting between the two branches at $\epsilon_{\bf k}$ and $\epsilon_{\bf k}+{\mathcal V}$, with the spectral weight controlled by $n_{\bf {k}+{Q}}$. This splitting is entirely driven by the interaction term in Eq.~(\ref{Hamiltonian}), in close analogy to the correlation-induced spectral reconstruction in the HK model.  The quantum critical point $p_{\rm cp}$ is therefore defined as the doping at which the correlation-induced pseudogap collapses.   This quantum critical point corresponds to the doping level at which the chemical potential $\mu$ reaches the lower edge of the upper branch, leading to the disappearance of the correlation-induced gap at the Fermi level ($\omega=0$). This defines $p_{\rm cp}$ as a correlation-driven transition between a pseudogap phase and a correlated metallic state within the exactly solvable structure of Eq.~(\ref{Hamiltonian}).}

Specifically, as shown in Fig.~\ref{fig:normal}(a), with increasing hole concentration, the system approaches a critical doping at $p=p_{\rm cp}$, where the chemical potential aligns with the lower boundary of the pseudogap. At this point, the gap collapses as dictated by the occupation-dependent structure of the spectral function in Eq.~(\ref{eq:A}), signaling a quantum phase transition between the pseudogap regime and a correlated metallic state. The disappearance of the pseudogap restores low-energy electronic states, giving rise to a pronounced enhancement of the density of states at the Fermi level, as shown in Fig.~\ref{fig:normal}(b). Upon further doping, the Fermi level intersects a weakly dispersive band in the antinodal region. This evolution culminates in a Lifshitz transition near $p\approx1.1p_{\rm cp}$, where the Fermi surface topology changes as the antinodal van Hove singularity is crossed by Fermi level.

In the heavily overdoped regime ($p>1.2p_{\rm cp}$), correlation-induced reconstruction becomes negligible. The low-energy spectrum closely resembles that of a noninteracting system, and a well-defined large Fermi surface is recovered. {
The effective mass, which reflects the density of states at the Fermi level, can be estimated as 
  ${m_{\rm eff}}/{m_e} \propto {N(E_F)}/{N_0(E_F)}$, where $N(E_F)=N(\omega=0)=\frac{1}{N}\sum_{\bf k} A(\bf k,\omega=0)$ is the interacting density of states at the Fermi level, with the spectral function $A(\mathbf{k},\omega)$ given by Eq.~(\ref{eq:A}), $N_0(E_F)=m_e/(2\pi \hbar^2)$ is the corresponding free-electron value.} Then, as shown in Fig.~\ref{fig:normal}(b), the calculated doping dependence of the Fermi-surface-averaged effective mass shows good agreement with experimental trends reported for cuprate superconductors~\cite{PhysRevResearch.3.043125,PhysRevB.106.195110,annurev:/content/journals/10.1146/annurev-conmatphys-030212-184305}.

\begin{figure}
    \centering 
  {\includegraphics[width=8.7cm]{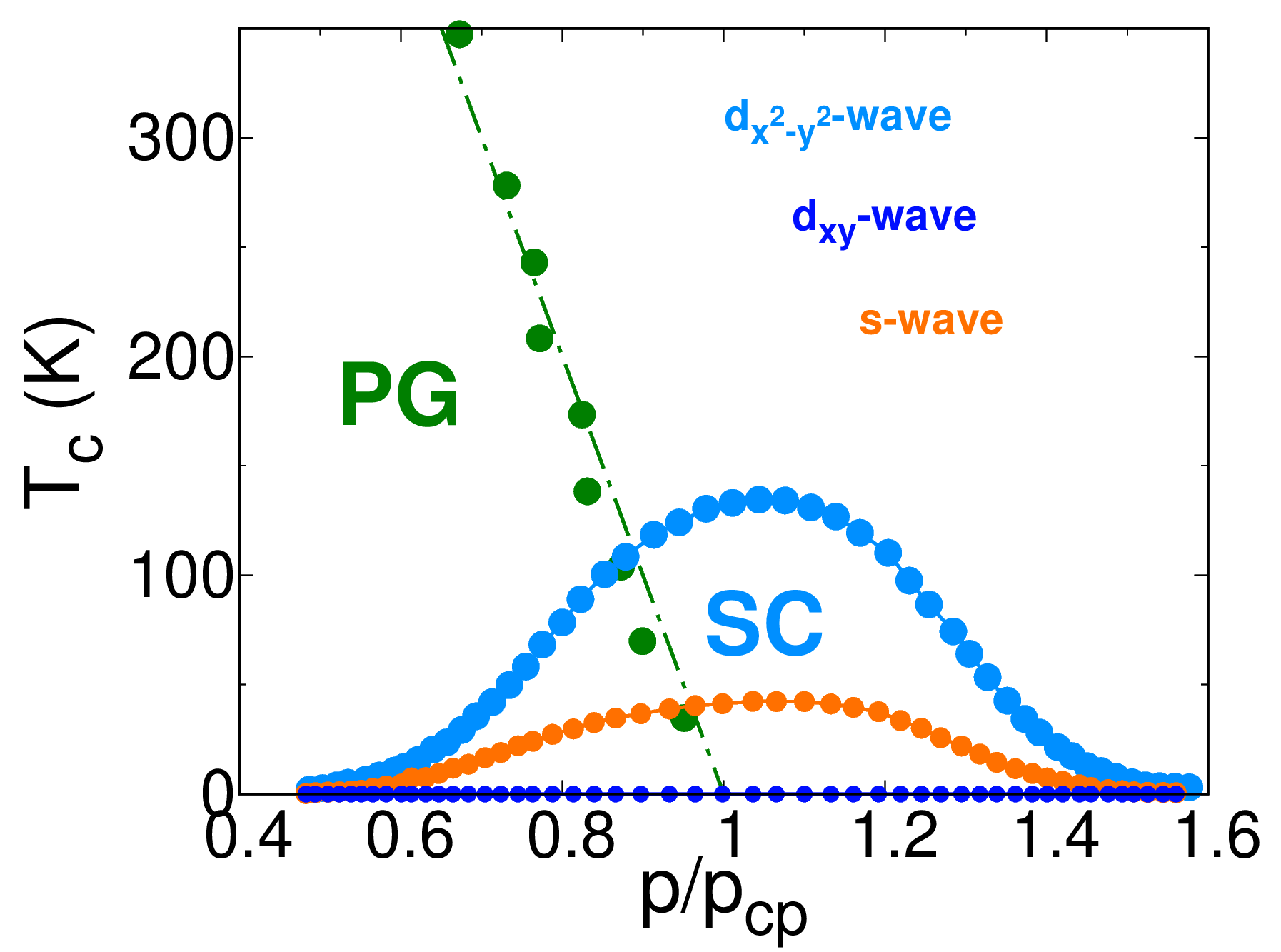}}
    \caption{The mean-field SC $T_c$ calculated by treating $\hat{H}$ exactly. We use a $256 \times 256$ $\bm{k}$ grid for the $\bm{k}$ summation in this simulation, and set the mean-field pairing potential in Eq.~(\ref{eq:gap}) as $J=0.55t$. {The $s$-wave results here  correspond only to the \emph{on-site} $s$-wave channel, and do not include extended $s$-wave components.}}
    \label{fig:y}
\end{figure}

  {The pseudogap temperature $T^*$ is defined as the temperature at which the suppression of spectral weight at the Fermi level ($\omega=0$) at the antinodal point ${\bf k}=(\pi,0)$ disappears, as encoded in the spectral function $A(\bf k,\omega)$ [Eq.~(\ref{eq:A})].} As shown in Fig.~\ref{fig:y}(a), the calculated characteristic pseudogap temperature $T^*$ decreases approximately linearly with increasing doping, consistent with experiments~\cite{keimer15from,dagotto94correlated,tom99the,armitage10progress,davis13concepts}.

\textcolor{black}{It is noted that neither $p_{\rm cp}$ nor $T^{*}$ corresponds to a conventional Landau phase transition characterized by the emergence of a local order parameter. Within the present exactly solvable framework, both $p_{\rm cp}$ and $T^{*}$ are defined directly from the single-particle spectral function. Specifically, $p_{\rm cp}$ is defined as the filling at which the correlation-induced pseudogap collapses at the Fermi level, corresponding to the point where the chemical potential reaches the lower edge of the upper spectral branch. Likewise, $T^{*}$ is defined as the temperature at which the antinodal suppression of spectral weight at the Fermi level disappears. Therefore, both quantities emerge naturally from the exact correlation-induced spectral reconstruction rather than a universally accepted symmetry-breaking continuous order parameter.}

{\sl SC Model.---}To investigate superconductivity emerging from the correlated normal state, we introduce pairing interactions on top of the effective Hamiltonian $\hat H$ and determine the   transition temperature $T_c$ within a self-consistent mean-field treatment. Specifically, {we consider an exchange interaction, 
\begin{equation}
\hat{H}_\mathrm{ex}
= \sum_{\langle i,j\rangle}{J_{ij}}\hat{\mathbf S}_i\cdot\hat{\mathbf S}_j
= \sum_{\langle i,j\rangle}\big[-\frac{1}{2}{J_{ij}}\hat b_{ij}^\dagger \hat b_{ij}
+\frac{1}{4}J_{ij}\hat n_i \hat n_j\big],
\label{eq:Hex_bij}
\end{equation}
where $\hat b_{ij}= \hat c_{i\downarrow}\hat c_{j\uparrow}-\hat c_{i\uparrow}\hat c_{j\downarrow}$ annihilates a spin-singlet pair on the bond $\langle i,j\rangle$, and $\hat n_i=\sum_\sigma \hat c_{i\sigma}^\dagger \hat c_{i\sigma}$.  This decomposition makes the global SU(2) spin-rotation invariance explicit and shows that antiferromagnetic exchange promotes spin-singlet pairing on the  bonds where $J_{ij}>0$. We focus on the pairing channel of the exchange interaction and drop the density–density term, which does not contribute to the spin-singlet pairing instability at the mean-field level (and can be absorbed into a redefinition of the normal-state dispersion/chemical potential).} The  SC Hamiltonian  
with a generalized pairing interaction in real space reads~\cite{li21superconductor}:
\begin{eqnarray}
    \hat{H}_\mathrm{SC}=\hat{H}-\frac{1}{2}\sum_{\langle i,j\rangle}{J_{ij}}\hat b_{ij}^\dagger \hat b_{ij},  
\end{eqnarray}
{In practice, we consider a generalized effective pairing interaction including on-site, nearest-neighbor, and next-nearest-neighbor components ($J_0$, $J_1$ and $J_2$). While $J_1$ and $J_2$ can be viewed as originating from antiferromagnetic exchange, the on-site term $J_0$ represents additional local pairing channels beyond pure antiferromagnetic exchange interaction. The corresponding momentum-space kernels for on-site, nearest-neighbor, and next-nearest-neighbor couplings on a square lattice are then written as $J^{(0)}_{\mathbf{k}\mathbf{q}} =J_0$, $J^{(1)}_{\mathbf{k}\mathbf{q}}= \tfrac{J_1}{2}[\cos(k_x-q_x)+\cos(k_y-q_y)]$, $J^{(2)}_{\mathbf{k}\mathbf{q}}=2J_2\cos(k_x-q_x)\cos(k_y-q_y)$, yielding 
\begin{align}
&J_{\bf kq}\!=\!\big[J_0\!+\!\frac{J_1}{4}(\cos{k_x}\!+\!\cos{k_y})(\cos{q_x}\!+\!\cos{q_y})\!+\!2J_2\cos{k_x}\nonumber\\
&\mbox{}\times\cos{k_y}\cos{q_x}\cos{q_y}\big]\!+\!\frac{J_1}{4}(\cos{k_x}\!-\!\cos{k_y})(\cos{q_x}\!-\!\cos{q_y})\nonumber\\
&\mbox{}+2J_2\sin{k_x}\sin{k_y}\sin{q_x}\sin{q_y}.
\end{align}
Projecting onto a given irreducible representation of the lattice symmetry, the kernel can be approximated in separable form as  $J_{\bf kq}=JC({\bf k})C({\bf q})$ where $C(\bm{k})$ encodes the pairing symmetry: $C(\bm{k})=\cos k_x-\cos k_y$ for $d_{x^2-y^2}$ channel, $C(\bm{k})=\sin k_x \sin k_y$ for $d_{xy}$ channel, and $C(\bm{k})\approx1$ for $s$-wave channel, as well as $C({\bf k})=(\cos{k_x}\!+\!\cos{k_y})$ or $C({\bf k})=\cos{k_x}\cos{k_y}$ for extended $s$-wave channel.}  Then, the SC Hamiltonian in momentum space takes the standard form: 
\begin{equation}
    \hat{H}_\mathrm{SC} = \hat{H} - \frac{J}{N} \sum_{\bm{k q}}C(\bm{k}) C(\bm{q}) \hat{c}^\dagger_{\bm{k}\uparrow} \hat{c}^\dagger_{-\bm{k}\downarrow} \hat{c}_{-\bm{q}\downarrow} \hat{c}_{\bm{q}\uparrow},
\end{equation}
where $N$ is the number of lattice sites. Applying a mean-field decoupling yields
\begin{equation}
    \hat{H}_\mathrm{SC}^\mathrm{MF} = \hat{H} + \sum_{\bm{k}} \Delta_{\bm{k}} (\hat{c}^\dagger_{\bm{k}\uparrow} \hat{c}^\dagger_{-\bm{k}\downarrow} + \mathrm{H.c.}), \label{eq:HMF}
\end{equation}
where the SC gap takes the form
$\Delta_{\bm{k}}=|\Delta|C(\bm{k})$. 
The gap amplitude $|\Delta|$ is determined self-consistently through
\begin{equation}
    |\Delta| = \frac{-J}{N} \sum_{\bm{k}} C(\bm{k}) \langle \hat{c}_{-\bm{k}\downarrow} \hat{c}_{\bm{k}\uparrow} \rangle. \label{eq:gap}
\end{equation}
Due to the presence of the AFM wave vector $\bm{Q}$ in the normal-state Hamiltonian, the mean-field problem naturally decomposes as $\hat{H}_\mathrm{SC}^\mathrm{MF} = \sum_{\bm{k} \in \mathrm{QBZ}} \sum_\sigma \hat{H}^\mathrm{MF}_{\bm{k} \sigma}$, where QBZ denotes one quarter of the Brillouin zone, with the remaining sectors generated by $\bm{k} \rightarrow -\bm{k}$ and $\bm{k} \rightarrow \bm{k} + \bm{Q}$), the corresponding block,
\begin{eqnarray}
    &&\hat{H}^\mathrm{MF}_{\bm{k} \sigma}= \epsilon_{\bm{k}} (\hat{n}_{\bm{k}\sigma} + \hat{n}_{-\bm{k},-{\sigma}}) + \epsilon_{\bm{k}+\bm{Q}} (\hat{n}_{\bm{k}+\bm{Q},-{\sigma}} + \hat{n}_{-\bm{k}-\bm{Q} \sigma}) \nonumber \\
    &&\mbox{}+{{\mathcal{V}}}(\hat{n}_{\bm{k}\sigma} \hat{n}_{\bm{k}+\bm{Q},-{\sigma}}\!+\!\hat{n}_{-\bm{k},-{\sigma}} \hat{n}_{-\bm{k}-\bm{Q}\sigma})\!+\!\big[C(\bm{k})\hat{c}^\dagger_{\bm{k}\sigma} \hat{c}^\dagger_{-\bm{k},-{\sigma}} \nonumber \\
   &&\mbox{}+C(\bm{k} + \bm{Q})\hat{c}^\dagger_{-\bm{k}-\bm{Q}\sigma} \hat{c}^\dagger_{\bm{k}+\bm{Q},-{\sigma}} + \mathrm{H.c.}\big]\sigma|\Delta|.
\end{eqnarray}
For each momentum $\bm{k}$, the mean-field Hamiltonian couples four single-particle states,  $(\bm{k},\sigma)$, $(-\bm{k},-{\sigma})$, $(\bm{k}+\bm{Q},-{\sigma})$, and $(-\bm{k}-\bm{Q},\sigma)$, giving rise to a $16$-dimensional many-body Hilbert space. The corresponding $16\times16$ matrix can therefore be diagonalized exactly for each $\bm{k}$, allowing the thermal averages entering Eq.~(\ref{eq:gap}) to be evaluated without further approximation.

\begin{figure}
    \centering
  {\includegraphics[width=8.6cm]{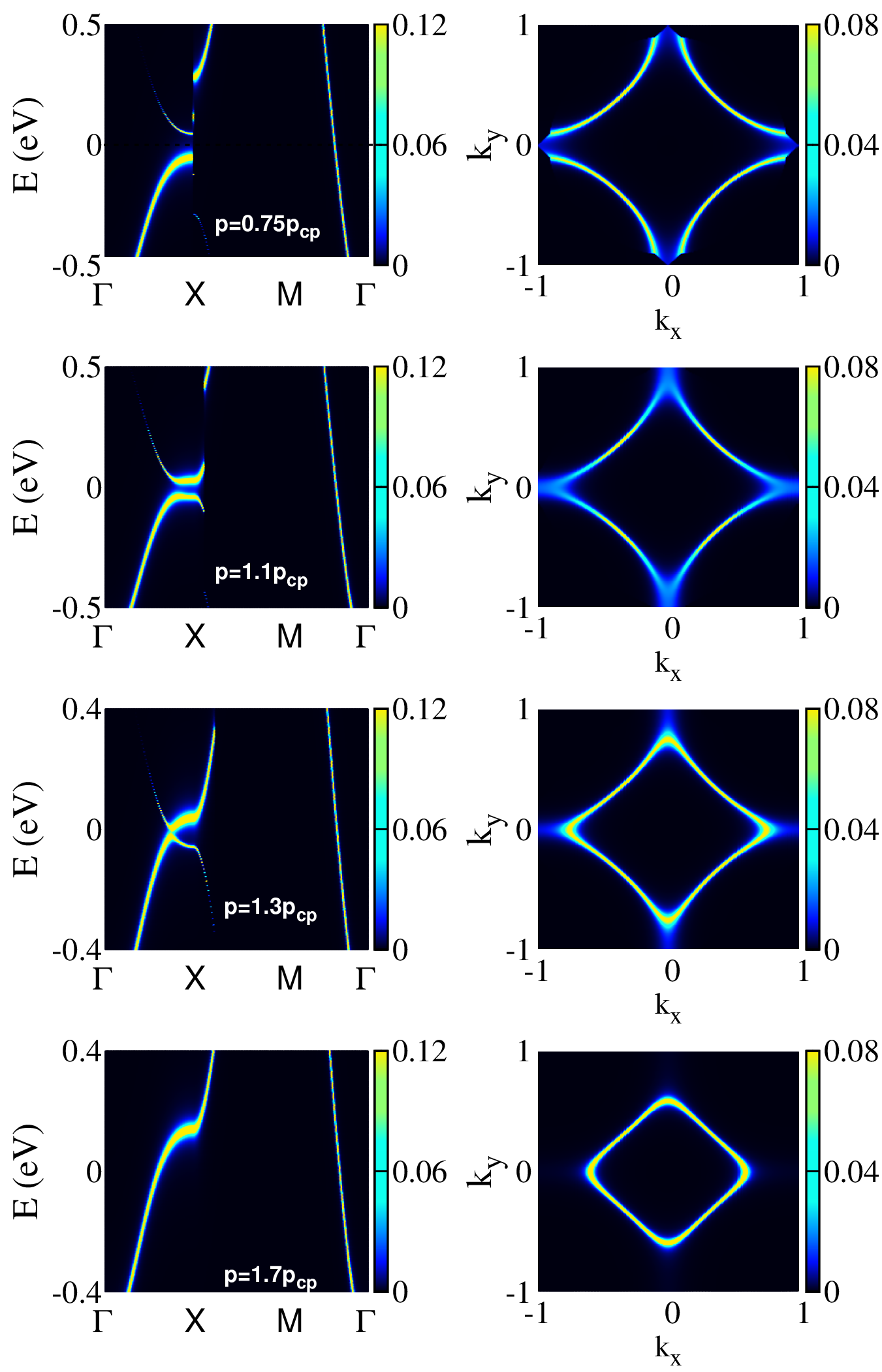}}
    \caption{Electronic structures of the SC-state phase at $k_BT\approx10^{-4}t$ for underdoped, quantum-critical point of pseudogap phase and overdoped cases.
        Left column: Single-particle excitation spectra along a high-symmetry path $\Gamma=(0, 0)$, $X=(\pi, 0)$, and $M=(\pi, \pi)$ in the Brillouin zone. Right column: Corresponding Fermi surfaces. We set the mean-field pairing potential in Eq.~(\ref{eq:gap}) as $J=0.55t$.}
    \label{fig:SC}
\end{figure}

We first 
discuss the spectral function in the SC state, as shown in Fig.~\ref{fig:SC}. Upon entering the SC phase, a well-defined energy gap opens around the Fermi level (left column of Fig.~\ref{fig:SC}), accompanied by the emergence of sharp and coherent Bogoliubov quasiparticle peaks (right column of Fig.~\ref{fig:SC}). The original normal-state dispersion is split into particle- and hole-like branches that are symmetric with respect to the Fermi energy (left column of Fig.~\ref{fig:SC}), reflecting the formation of phase-coherent Cooper pairs. The SC gap exhibits a pronounced momentum dependence, vanishing along the nodal directions while reaching its maximum near the antinodal regions, consistent with a $d_{x^2-y^2}$-wave pairing symmetry. This is accompanied by a pronounced transfer of low-energy spectral weight away from the antinodal momenta toward the nodal directions (right column of Fig.~\ref{fig:SC}), signaling a \emph{coherent} reorganization of the single-particle spectrum in the SC state. This behavior is qualitatively distinct from that of the pseudogap regime shown in Fig.~\ref{fig:normal}, where the spectral weight is strongly suppressed or completely eliminated in a \emph{highly localized} region of momentum space near the antinodal point $(\pi,0)$. In contrast to such a momentum-localized and incoherent suppression, the SC gap reorganizes the low-energy spectrum coherently over the entire Fermi surface, leaving gapless excitations only along the nodal directions (left column of Fig.~\ref{fig:SC}). These features are qualitatively consistent with experimental observations in cuprate superconductors, including the emergence of coherent Bogoliubov quasiparticles, the momentum-dependent 
$d_{x^2-y^2}$-wave gap, and the transfer of low-energy spectral weight toward the nodal directions observed in angle-resolved photoemission spectroscopy~\cite{PhysRevLett.90.217002,Vishik2012,RevModPhys.75.473,Ding1996,loeser1996excitation,norman1998,Hashimoto2014,Kanigel2006,Lee2007}.

The SC transition temperature $T_c$ is determined as the temperature at which the self-consistent gap amplitude $|\Delta|$ vanishes. Figure~\ref{fig:y} shows the mean-field $T_c$ as a function of hole doping $p$. To enable a direct comparison between different pairing symmetries, we assign the same pairing strength $J$ to the $s$-wave, $d_{x^2-y^2}$-wave, and $d_{xy}$-wave channels. This allows us to assess their relative stability under identical interaction conditions. The resulting temperature–doping phase diagram demonstrates that the correlated normal-state electronic structure strongly enhances $d_{x^2-y^2}$-wave superconductivity, completely suppresses the $d_{xy}$-wave instability, and significantly disfavors the $s$-wave channel. This behavior contrasts sharply with the HK model, which instead favors $s$-wave pairing.

Notably, the $d_{x^2-y^2}$-wave SC transition temperature displays a pronounced dome-shaped dependence on hole doping. This nonmonotonic behavior is rooted in the doping evolution of the correlated normal-state electronic structure.  The SC dome emerges only beyond a finite critical doping $p\approx0.43p_{\rm cp}$. For lower dopings, superconductivity is suppressed by the strong-correlation-induced collapse/depletion  of low-energy spectral weight of the single-particle spectrum encoded in Eq.~(\ref{eq:A}). Above this critical doping in the underdoped regime, increasing doping progressively restores spectral weight in the antinodal region, where the pseudogap is most prominent. The weakening of the pseudogap enhances low-energy quasiparticle coherence and, in turn, strengthens SC pairing.
 
 As doping approaches $p=p_{\rm cp}$, the collapse of the pseudogap leads to a strong enhancement of the correlated normal-state density of states at the Fermi level [Fig.~\ref{fig:normal}(b)]. The rapid recovery of low-energy spectral weight creates an optimal environment for superconductivity. Interestingly, however, the maximum $T_c$ does not occur exactly at the pseudogap closing point. Instead, within the present tight-binding parameter set, the optimal SC doping is shifted to slightly higher values and coincides with a Lifshitz transition near $p\approx1.1p_{\rm cp}$, where the Fermi level crosses the antinodal van Hove singularity. This shift highlights the combined role of correlation-driven spectral reconstruction and Fermi-surface topology in determining SC stability.  Upon further doping into the overdoped regime, correlation effects become progressively weaker, and the electronic structure approaches that of an uncorrelated metal. Correspondingly, the SC transition temperature decreases monotonically with increasing doping, in agreement with calculations based on uncorrelated normal states~\cite{phillips20exact,li21superconductor}.  As a consequence, the SC dome observed here is entirely driven by electronic correlations in the normal state.

{Consequently, the pseudogap correlations and superconductivity are competing.  The pseudogap, arising from the correlation term in Eq.~(\ref{Hamiltonian}), leads to a suppression of spectral weight at the Fermi level, particularly in the antinodal region, which effectively reduces the low-energy density of states available for pairing. As a result, superconductivity is suppressed in the underdoped regime. In contrast, in the absence of such correlations, the density of states remains finite and $T_c$ would increase monotonically with decreasing $p/p_{\rm cp}$, without forming a dome~\cite{li21superconductor}.  Therefore, the SC dome emerges from the competition between the correlation-induced depletion of low-energy spectral weight (pseudogap) and the SC pairing. }

{\sl Discussion.---}The present work reported the central role of correlation-induced spectral reconstruction in shaping superconductivity in cuprates. Our analysis reveals the emergence of a correlation-driven SC dome, the energetic dominance of the $d_{x^2-y^2}$-wave pairing channel, and a close connection between pseudogap correlations and SC stability. Importantly, the pseudogap and the SC gap exert fundamentally different effects on the single-particle spectrum: while the pseudogap leads to an incoherent suppression and partial removal of low-energy spectral weight, predominantly near the antinodal regions, the SC gap redistributes spectral weight coherently, sharpening quasiparticle peaks and transferring spectral weight toward the nodal directions without eliminating it.

The simplified exactly solvable correlated model employed in this work cannot capture all microscopic aspects of strong correlations in cuprates. Nevertheless, it provides a controlled and analytically transparent framework that is sufficiently rich to describe the essential normal-state physics and its nontrivial interplay with superconductivity. Within this framework, the emergence of a SC dome and the stabilization of $d$-wave pairing arise as robust consequences of correlation-induced spectral reconstruction. By contrast, in the absence of such correlations, the SC transition temperature would evolve monotonically with doping, and no dome structure would be expected. As such, our results offer a controlled route toward isolating the key mechanisms underlying the SC dome and clarifying the role of pseudogap correlations in high-$T_c$ cuprates.

We note that the quantitative doping location of $p_{\rm cp}$ discussed here depends on the specific choice of parameters used to model the noninteracting band structure. Its specific value, and therefore the overall alignment of the phase diagram along the doping axis, can be tuned by adjusting the underlying band parameters. We expect such quantitative shifts should not affect the qualitative conclusions of this work.

{\sl Acknowledgments.---}Z.W.P. acknowledges the financial support by the National Natural Science Foundation of China under
Grants No.~12375292 and No.~12005221.

\bibliography{sc-refs}

\appendix

\end{document}